\documentclass[preprint,prd,amsmath,amssymb,aps,nofootinbib]{revtex4-1}

\usepackage{graphicx}
\usepackage{siunitx}
\usepackage{bm}
\usepackage{natbib}
\graphicspath{{figures/}}
\usepackage{threeparttable}

\def\apj{Astrophys.\ J.\ }

\begin{document}

\title{A New Perspective on the Diffuse Gamma-Ray Emission Excess}
\author{En-Sheng Chen$^{a,b}$}
  \email{chenes@ihep.ac.cn}
\author{Kun Fang$^{a}$}
  \email{fangkun@ihep.ac.cn}
\author{Xiao-Jun Bi$^{a,b}$}
  \email{bixj@ihep.ac.cn}
\affiliation{
$^a$Key Laboratory of Particle Astrophysics, Institute of High Energy
Physics, Chinese Academy of Sciences, Beijing 100049, China \\
$^b$University of Chinese Academy of Sciences, Beijing 100049, China\\
}


\date{\today}

\begin{abstract}
The Large High-Altitude Air Shower Observatory (LHAASO) recently published measurements of diffuse Galactic gamma-ray emission (DGE) in the 10-1000 TeV energy range. The measured DGE flux is significantly higher than the expectation from hadronic interactions between cosmic rays (CRs) and the interstellar medium. This excess has been proposed to originate from unknown extended sources produced by electron radiation, such as pulsar wind nebulae or pulsar halos (PWNe/halos). In this study, we propose a new perspective to explain the DGE excess observed by LHAASO. The masking regions used in the LHAASO DGE measurement may not fully encompass the extended signals of PWNe/halos. By employing a two-zone diffusion model for electrons around pulsars, we find that the DGE excess in most regions of the Galactic plane can be well explained by the signal leakage model under certain parameters. Our results indicate that the signal leakage from known sources and contributions from unresolved sources should be considered complementary in explaining the DGE excess.
\end{abstract}

\maketitle

\section{Introduction}          \label{sec:intro}
The diffuse Galactic gamma-ray emission (DGE) is crucial for studying the origin and propagation of cosmic rays (CRs). Traditionally, the DGE from the Galactic plane is believed to be dominantly generated by the interactions between the propagating CRs and interstellar medium (ISM)  \cite{Strong:2007nh,Fermi-LAT:2012edv}. However, at the TeV-PeV energy region, the DGE predicted by the CR-ISM interaction model is significantly lower than those observed by Milagro, ARGO-YBJ, and the Tibet AS+MD array \cite{Milagro:2005xqq,ARGO-YBJ:2015cpa,TibetASgamma:2021tpz}, which is known as the TeV DGE excess \cite{Prodanovic:2006bq,Linden:2017blp}. 

Recently, the Large High-Altitude Air Shower Observatory (LHAASO) published the gamma-ray source catalog of the TeV-PeV energy band \cite{LHAASO:2023rpg} and the measurement of DGE in the 10-1000 TeV energy range by masking the source regions \cite{LHAASO:2023gne}. The measured DGE flux in the inner Galaxy region is about 3 times higher than the expectation from the CR-ISM interactions and about 2 times higher than that expected in the outer Galaxy region \cite{Zhang:2023ajh}. On the other hand, neutrinos are produced alongside gamma rays during the CR-ISM interactions. The IceCube neutrino telescope has measured the high-energy neutrino flux from the Galactic plane in the energy range of 1-100~TeV \cite{IceCube:2023ame}. Subtracting possible contribution of point sources from the total neutrino measurement and applying the masking method of LHAASO, Ref.~\cite{Yan:2023hpt} argued that the gamma-ray flux associated with the neutrino flux is consistent with that predicted by the CR-ISM interactions. This result supports that the TeV DGE excess is mainly contributed by leptonic processes\footnote{Due to the model-dependent nature of the neutrino flux measurement and the associated uncertainties, we cannot rule out the possibility that the DGE excess includes a minor hadronic component.}.

In the LHAASO DGE measurement, each source region is masked with a 2.5 times Gaussian width of the source. For point sources and Gaussian-like sources, this method can effectively remove the source contamination to the DGE. However, the morphology of pulsar wind nebulae (PWNe) and pulsar halos can be more extended than the Gaussian distribution. Spatially dependent transport of electrons and positrons\footnote{\textit{Electrons} will denote both electrons and positrons hereafter if not specified.}, such as two-zone diffusion, is suggested for these systems \cite{2020ApJ...889...30L,Fang:2023xla}, indicating that there could be more gamma-ray signals than predicted by the Gaussian profile at large distances from the source. Many gamma-ray sources in the first LHAASO catalog are associated with pulsars, most of which could be PWNe or pulsar halos. Thus, the potential contamination of DGE by these sources is worth considering. We refer to all sources associated with pulsars as LHAASO PWNe/halos in this paper. 

More intriguingly, the Galactic longitude profile of the DGE excess measured by LHAASO exhibits a correlation with some known gamma-ray sources that have large extensions, such as those in the Geminga and Cygnus regions. The significant increase in DGE around Geminga is likely due to signal leakage from the Geminga pulsar halo. For the Cygnus region, the measurement by LHAASO indicates that the bubble extends to at least $10^\circ$ \cite{LHAASO:2023uhj}, while the size of the masked region at that location is $6^\circ$. Therefore, signals beyond $6^\circ$ can contribute to the measurement of DGE.

In this work, we introduce a novel approach by evaluating the contribution of leakage signals from LHAASO PWNe/halos to the TeV DGE excess based on the two-zone diffusion model of electrons. This perspective is entirely distinct from the previous attribution of the DGE excess to unresolved extended gamma-ray sources \cite{Linden:2017blp,Dekker:2023six,Yan:2023hpt}. Additionally, we give an estimate of the signal leakage from the Cygnus bubble. In Section \ref{sec:Methods}, we introduce the two-zone diffusion model, as well as the measurements of LHAASO PWNe/halos in the first LHAASO catalog and the related pulsar characteristics. We adopt the two-zone diffusion model to fit the LHAASO measurements and then predict the signal leakage. Section \ref{sec:result} presents our results, including the Galactic longitude profile and gamma-ray energy spectrum contributed by the leakage signals. Section \ref{sec:discussion} provides some extended discussions of the results, followed by a summary and outlook in Section \ref{sec:conclusion}.

\section{Methods} \label{sec:Methods}
The first LHAASO catalog of gamma-ray sources includes 90 sources, 35 of which are associated with pulsars. We assume these sources to be PWNe or pulsar halos. In the LHAASO catalog paper, their morphology is uniformly described using a Gaussian template \cite{LHAASO:2023rpg}. While this approach effectively captures the signals near the pulsars, it may not accurately describe the gamma-ray signals at larger angular distances considering the possible escape of the parent electrons from the central zone. In this study, we adopt a two-zone diffusion model to describe the gamma-ray surface brightness of these sources, ensuring that the central morphology of each source is in agreement with the LHAASO measurements. The leakage flux can then be estimated by extrapolating from the two-zone diffusion model.

\subsection{Two-zone model} \label{subsec:2zone}
We assume that the electron propagation for the LHAASO PWNe/halos can be described by the diffusion-loss equation: 
\begin{equation}
\frac{\partial N\left(E_e, \boldsymbol{r}, t\right)}{\partial t}=\nabla \cdot\left[D\left(E_e\right) \nabla N\left(E_e, \boldsymbol{r}, t\right)\right]+\frac{\partial\left[b\left(E_e\right) N\left(E_e, \boldsymbol{r}, t\right)\right]}{\partial E_e}+Q\left(E_e, \boldsymbol{r}, t\right)\,,
\label{eq:propagation}
\end{equation}
where $N$ is the differential electron number density at electron energy $E_e$, position $\boldsymbol{r}$ and time $t$. $D$ is the diffusion coefficient, $b\equiv |dE_e/dt|$ is the energy-loss rate due to electromagnetic radiation, and $Q$ is the source term.

The source is assumed to be point-like, and the electron injection rate is assumed to follow the time profile of the pulsar spin-down luminosity as as $\propto\left(1+t / \tau \right)^{-2}$, where the spin-down time scale is set to be $\tau=10~\rm{kyr}$. The electron injection spectrum for each source is described by a power-law form. Hence, the source injection function is expressed as
\begin{equation} \label{eq:inj}
    Q\left(E_e, \boldsymbol{r}, t\right)= \begin{cases} q_{0}E^{-\alpha} \delta\left(\boldsymbol{r}-\boldsymbol{r}_p\right)\left[\left(t_p+\tau\right) /\left(t+\tau\right)\right]^2, & t \geq 0 \\ 0, & t<0\end{cases}\,,
\end{equation}
where $\boldsymbol{r}_p$ is the position of pulsar, decided by the pulsar distance $d$ listed in Table~\ref{tab:pulsar}. $t_{p}$ is the age of pulsar, and $t=0$ corresponds to the birth time of pulsar.

For the two-zone model, the diffusion coefficient takes the form of
\begin{equation}
    D\left(E_e, \boldsymbol{r}\right)= \begin{cases}D_1(E_e/100~\mathrm{TeV})^{\delta}, & \left|\boldsymbol{r}-\boldsymbol{r}_p\right|<r_{\star} \\ D_2(E_e/100~\mathrm{TeV})^{\delta}, & \left|\boldsymbol{r}-\boldsymbol{r}_p\right| \geq r_{\star}\end{cases}
\end{equation}
where $r_{\star}$ is the size of slow-diffusion zone, $D_1$ is the suppressed diffusion coefficient near the source, and $D_2$ is the typical diffusion coefficient of the Galaxy \cite{Ma:2022iji}. Both $D_1$ and $D_2$ are normalized at 100~TeV. The energy slope of the diffusion coefficient is assumed to be $\delta=1/3$ as suggested by Kolmogorov’s theory \cite{Kolmogorov1941}. Unless specified, we take $D_1=4.5\times10^{27}$~cm$^2$~s$^{-1}$ as inferred from the surface brightness profile of the Geminga halo \cite{Abeysekara2017} and the size of the slow-diffusion zone to be $r_{\star}=25$ pc.

For high-energy electrons ($E_e\gg1$~GeV), energy losses are dominated by synchrotron radiation and inverse Compton scattering (ICS) \cite{Fang:2020dmi}, which is denoted as $b(E_e)$ in equation \ref{eq:propagation}. We use a $3$~$\mu$G magnetic field to calculate the energy loss due to synchrotron radiation. For ICS, the Galactic interstellar radiation field (ISRF) is composed of the cosmic microwave background, infrared emission, and optical emission. These components are described by gray body distributions with temperatures of $2.7$~K, $20$~K, and $5000$~K, and energy densities of $0.26$~eV~cm$^{-3}$, $0.3$~eV~cm$^{-3}$, and $0.3$~eV~cm$^{-3}$, respectively. 

We use a finite volume numerical method to solve the equation \ref{eq:propagation} to obtain the electron density $N(E_e, \boldsymbol{r}, t)$ \cite{Fang:2018qco}. Electrons produce gamma rays through ICS with the Galactic ISRF \cite{Blumenthal:1970gc}. We then perform line-of-sight integration to obtain the gamma-ray surface brightness $S(\theta, E_{\gamma})$ around pulsars, where $\theta$ is the angular distance from pulsars.

\begin{table}[htbp]
\centering
\caption[]{1LHAASO sources associated with pulsars.}
\label{tab:pulsar}
\resizebox{\textwidth}{!}{
\begin{tabular}{cccccccccc}
 \hline \hline
 Name & RA($^\circ$) & DEC($^\circ$) & $N_{0}\rm{(10^{-16}cm^{-2}s^{-1}TeV^{-1})}$ & $\Gamma$ & $\sigma(^\circ)$ & Associated Pulsar & $d$ (kpc) & $t_p$(kyr) & $\Dot{E}$(erg/s) \\
 \hline
 1LHAASO J0007+7303u & $1.91$ & $73.07$ & $3.41 \pm 0.27$ & $3.4 \pm 0.12$ & $0.17 \pm 0.03$ & PSR J0007+7303 & $1.4$ & $14$ & $4.50 \times 10^{35}$ \\
 1LHAASO J0216+4237u & $34.1$ & $42.63$ & $0.18 \pm 0.03$ & $2.58 \pm 0.17$ & $0.13^{\dag}$ & PSR J0218+4232 & $3.15$ & $476000$ & $2.40 \times 10^{35}$ \\
 1LHAASO J0249+6022 & $42.39$ & $60.37$ & $0.93 \pm 0.09$ & $3.82 \pm 0.18$ & $0.38 \pm 0.08$ & PSR J0248+6021 & $2$ & $62$ & $2.10 \times 10^{35}$ \\
 1LHAASO J0359+5406 & $59.78$ & $54.1$ & $0.85 \pm 0.06$ & $3.84 \pm 0.15$ & $0.3 \pm 0.04$ & PSR J0359+5414 & $3.8^{*}$ & $75$ & $1.30 \times 10^{36}$ \\
 1LHAASO J0534+2200u & $83.61$ & $22.04$ & $6.23 \pm 0.1$ & $3.19 \pm 0.03$ & $0.06^{\dag} $ & PSR J0534+2200 & $2$ & $1$ & $4.50 \times 10^{38}$ \\
 1LHAASO J0542+2311u & $85.71$ & $23.2$ & $2.93 \pm 0.12$ & $3.74 \pm 0.09$ & $0.98 \pm 0.05$ & PSR J0543+2329 & $1.56^{*}$ & $253$ & $4.10 \times 10^{34}$ \\
 1LHAASO J0622+3754 & $95.5$ & $37.9$ & $1.42 \pm 0.07$ & $3.68 \pm 0.1$ & $0.46 \pm 0.03$ & PSR J0622+3749 & $1.6$ & $208$ & $2.70 \times 10^{34}$ \\
 1LHAASO J0631+1040 & $97.77$ & $10.67$ & $0.54 \pm 0.06$ & $3.33 \pm 0.16$ & $0.3^{\dag}$ & PSR J0631+1037 & $2.1$ & $44$ & $1.70 \times 10^{35}$ \\
 1LHAASO J0634+1741u & $98.57$ & $17.69$ & $4.42 \pm 0.15$ & $3.69 \pm 0.06$ & $0.89 \pm 0.04$ & PSR J0633+1746 & $0.19$ & $342$ & $3.30 \times 10^{34}$ \\
 1LHAASO J0635+0619 & $98.76$ & $6.33$ & $0.94 \pm 0.1$ & $3.67 \pm 0.18$ & $0.6 \pm 0.07$ & PSR J0633+0632 & $1.35$ & $59$ & $1.20 \times 10^{35}$ \\
 1LHAASO J1740+0948u & $265.03$ & $9.81$ & $0.41 \pm 0.04$ & $3.13 \pm 0.15$ & $0.11^{\dag} $ & PSR J1740+1000 & $1.23$ & $114$ & $2.30 \times 10^{35}$ \\
 1LHAASO J1809-1918u & $272.38$ & $-19.3$ & $9.46 \pm 1.27$ & $3.51 \pm 0.26$ & $0.22^{\dag} $ & PSR J1809-1917 & $3.27$ & $51$ & $1.80 \times 10^{36}$ \\
 1LHAASO J1813-1245 & $273.36$ & $-12.75$ & $1.42 \pm 0.27$ & $3.66 \pm 0.34$ & $0.31^{\dag} $ & PSR J1813-1245 & $2.63$ & $43$ & $6.20 \times 10^{36}$ \\
 1LHAASO J1825-1256u & $276.44$ & $-12.94$ & $5.08 \pm 0.42$ & $3.33 \pm 0.13$ & $0.2^{\dag} $ & PSR J1826-1256 & $1.55$ & $14$ & $3.60 \times 10^{36}$ \\
 1LHAASO J1825-1337u & $276.45$ & $-13.63$ & $10.1 \pm 0.61$ & $3.28 \pm 0.09$ & $0.18^{\dag} $ & PSR J1826-1334 & $3.61$ & $21$ & $2.80 \times 10^{36}$ \\
 1LHAASO J1837-0654u & $279.31$ & $-6.86$ & $3.06 \pm 0.21$ & $3.7 \pm 0.12$ & $0.33 \pm 0.04$ & PSR J1838-0655 & $6.6$ & $23$ & $5.60 \times 10^{36}$ \\
 1LHAASO J1839-0548u & $279.79$ & $-5.81$ & $3.03 \pm 0.2$ & $3.24 \pm 0.09$ & $0.22 \pm 0.02$ & PSR J1838-0537 & $2.3^{*}$ & $5$ & $6.00 \times 10^{36}$ \\
 1LHAASO J1848-0001u & $282.19$ & $-0.02$ & $1.64 \pm 0.1$ & $2.75 \pm 0.07$ & $0.09^{\dag} $ & PSR J1849-0001 & $1.9^{*}$ & $43$ & $9.80 \times 10^{36}$ \\
 1LHAASO J1857+0245 & $284.37$ & $2.75$ & $<0.32$ & -  & $0.24 \pm 0.04$ & PSR J1856+0245 & $6.32$ & $21$ & $4.60 \times 10^{36}$ \\
 1LHAASO J1906+0712 & $286.56$ & $7.2$ & $<0.19$ & -  & $0.21 \pm 0.05$ & PSR J1906+0722 & $1.73^{*}$ & $49$ & $1.00 \times 10^{36}$ \\
 1LHAASO J1908+0615u & $287.05$ & $6.26$ & $6.86 \pm 0.16$ & $2.82 \pm 0.03$ & $0.36 \pm 0.01$ & PSR J1907+0602 & $2.37$ & $20$ & $2.80 \times 10^{36}$ \\
 1LHAASO J1912+1014u & $288.38$ & $10.5$ & $1.52 \pm 0.1$ & $3.26 \pm 0.11$ & $0.5 \pm 0.04$ & PSR J1913+1011 & $4.61$ & $169$ & $2.90 \times 10^{36}$ \\
 1LHAASO J1914+1150u & $288.73$ & $11.84$ & $0.79 \pm 0.06$ & $3.41 \pm 0.13$ & $0.21 \pm 0.04$ & PSR J1915+1150 & $14.01$ & $116$ & $5.40 \times 10^{35}$ \\
 1LHAASO J1928+1746u & $292.17$ & $17.89$ & $0.72 \pm 0.07$ & $3.1 \pm 0.12$ & $0.16^{\dag} $ & PSR J1928+1746 & $4.34$ & $83$ & $1.60 \times 10^{36}$ \\
 1LHAASO J1929+1846u & $292.04$ & $18.97$ & $0.64 \pm 0.06$ & $3.11 \pm 0.12$ & $0.21^{\dag} $ & PSR J1930+1852 & $7$ & $3$ & $1.20 \times 10^{37}$ \\
 1LHAASO J1954+2836u & $298.55$ & $28.6$ & $0.42 \pm 0.05$ & $2.92 \pm 0.14$ & $0.12^{\dag} $ & PSR J1954+2836 & $1.96$ & $69$ & $1.10 \times 10^{36}$ \\
 1LHAASO J1954+3253 & $298.63$ & $32.88$ & $<0.04$ & - & $0.17 \pm 0.03$ & PSR J1952+3252 & $3$ & $107$ & $3.70 \times 10^{36}$ \\
 1LHAASO J1959+2846u & $299.78$ & $28.78$ & $0.84 \pm 0.07$ & $2.9 \pm 0.1$ & $0.29 \pm 0.03$ & PSR J1958+2845 & $1.95$ & $22$ & $3.40 \times 10^{35}$ \\
 1LHAASO J2005+3415 & $301.81$ & $33.87$ & $0.56 \pm 0.05$ & $3.79 \pm 0.21$ & $0.33 \pm 0.05$ & PSR J2004+3429 & $10.78$ & $18$ & $5.80 \times 10^{35}$ \\
 1LHAASO J2005+3050 & $301.45$ & $30.85$ & $0.46 \pm 0.05$ & $3.62 \pm 0.21$ & $0.27 \pm 0.05$ & PSR J2006+3102 & $6.04$ & $104$ & $2.20 \times 10^{35}$ \\
 1LHAASO J2020+3649u & $305.23$ & $36.82$ & $2.29 \pm 0.09$ & $3.31 \pm 0.06$ & $0.12 \pm 0.02$ & PSR J2021+3651 & $1.8$ & $17$ & $3.40 \times 10^{36}$ \\
 1LHAASO J2028+3352 & $307.21$ & $33.88$ & $1.61 \pm 0.19$ & $3.38 \pm 0.19$ & $1.7 \pm 0.23$ & PSR J2028+3332 & $0.91^{*}$ & $576$ & $3.50 \times 10^{34}$ \\
 1LHAASO J2031+4127u & $307.95$ & $41.46$ & $2.56 \pm 0.08$ & $3.45 \pm 0.06$ & $0.22 \pm 0.01$ & PSR J2032+4127 & $1.33$ & $201$ & $1.50 \times 10^{35}$ \\
 1LHAASO J2228+6100u & $337.01$ & $61$ & $4.76 \pm 0.14$ & $2.95 \pm 0.04$ & $0.35 \pm 0.01$ & PSR J2229+6114 & $3$ & $10$ & $2.20 \times 10^{37}$ \\
 1LHAASO J2238+5900 & $339.54$ & $59$ & $2.03 \pm 0.12$ & $3.55 \pm 0.09$ & $0.43 \pm 0.03$ & PSR J2238+5903 & $2.83$ & $27$ & $8.90 \times 10^{35}$ \\
\hline \hline
\end{tabular}
}
\begin{tablenotes}
\footnotesize
\item[1] [1] The power-law shape is defined by 
$dN/dE_\gamma = N_0(E_\gamma/50~\mathrm{TeV})^{-\Gamma}$.
\item[2] [2] The pulsar distances marked with $*$ are ``pseudo distances" (see the text for the definition).
\item[3] [3] The Gaussian widths marked with $\dag$ are the $95\%$ statistical upper limits.
\item[4] [4] PSR is the abbreviation for \textit{pulsar}.
\end{tablenotes}
\end{table}

\subsection{LHAASO PWNe/halos}
Among these 35 LHAASO PWNe/halos, 32 have been significantly detected by LHAASO-KM2A and are utilized in this study except for the Crab nebula, as it is a very young source with few expected escaping electrons. In the first LHAASO catalog, their morphologies are all described by the Gaussian template, where $\sigma$ represents the Gaussian width or its the $95\%$ upper limit. Their energy spectra follow a power-law shape as $dN/dE_\gamma = N_0(E_\gamma/50~\mathrm{TeV})^{-\Gamma}$. The gamma-ray spectral parameters of the LHAASO sources and parameters of the associated pulsars are listed in Table~\ref{tab:pulsar}. For most cases, the pulsar distance, characteristic age, and spin-down luminosity are taken from the ATNF catalog \cite{Manchester:2004bp}. When the pulsar distance is not available in the ATNF catalog, we use the ``pseudo distance" derived from the empirical relation between the pulsar gamma-ray luminosity and total spin-down power \cite{Fermi-LAT:2009orv}. Specifically, PSR J1849-0001 currently lacks gamma-ray pulse measurements. We assume its distance to be the same as that of PSR J1954+2836, as these two pulsars have similar ages and Gaussian widths.

\subsection{Fitting method} \label{subsec:fit}
\begin{figure}
   \centering
   \includegraphics[width=0.6\textwidth, angle=0]{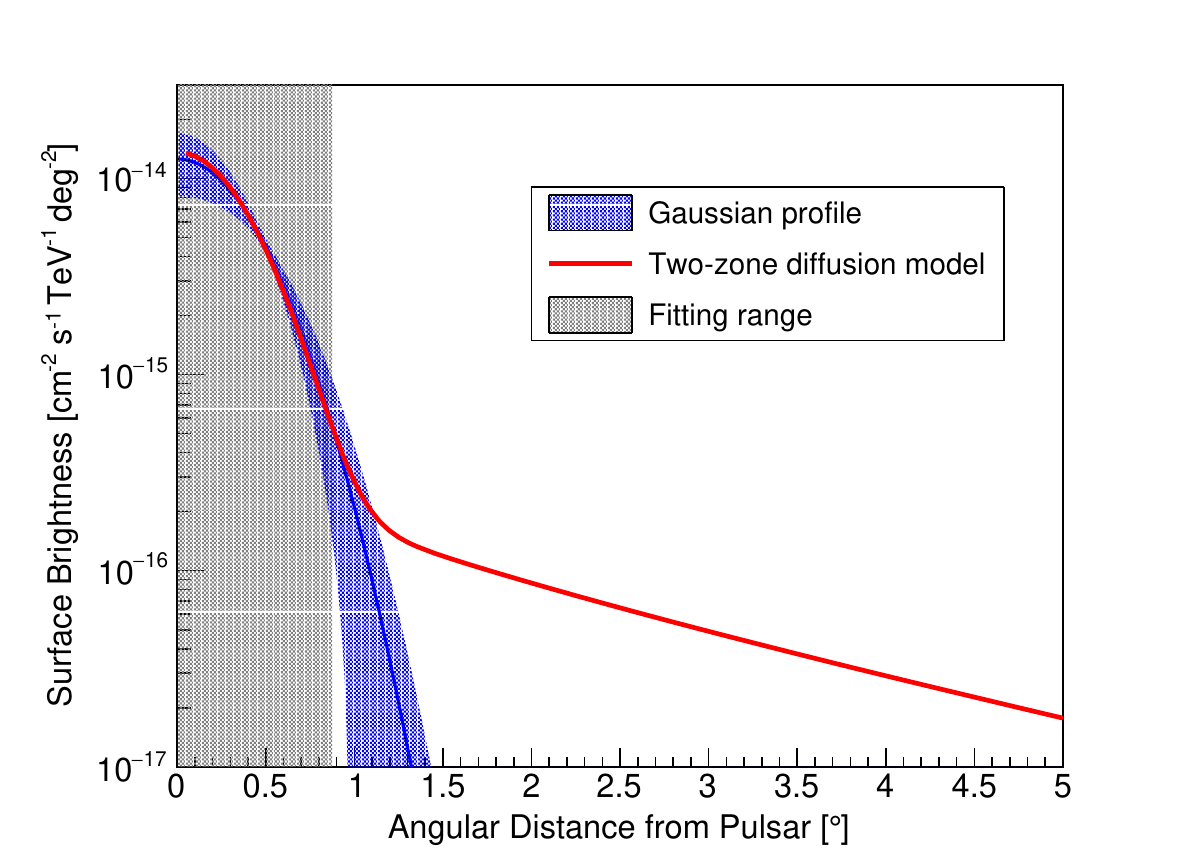}
   \caption{The surface brightness profile of LHAASO J1825-1337u with $E_{\gamma}=32$~TeV. The blue line represents the prediction of the Gaussian profile, with corresponding shaded area indicating the error band, assuming the parameter errors are independent of each other. The red line represents the expectation from the two-zone diffusion model. The gray shaded area indicates the fitting range.}
   \label{fig:Gaussian profile}
\end{figure}

Based on the LHAASO measurements of the morphology and energy spectrum for each source, we can obtain the Gaussian profile at each energy point between 10 TeV and 1 PeV. Taking LHAASO J1825-1337u as an example, the Figure \ref{fig:Gaussian profile} shows the Gaussian morphology at $E_{\gamma}=32$ TeV, convolved with the point spread function (PSF) of LHAASO-KM2A, where the PSF width is $0.3^\circ$. The blue shaded band represents the statistical error, assuming that the errors of each parameter are independent. In order to accurately reproduce the gamma-ray profile within the central zone and minimize the influence of uncertainties at large angular distances, we fit the two-zone diffusion model to the Gaussian template within 2.5$\sigma$ range, employing a chi-square fitting method as:
\begin{equation}
    \chi^{2}=\sum_{i}^{n} \left(\frac{S(\theta, E_{\gamma})-S_{G}(\theta, E_{\gamma})}{\sigma_{G}^i} \right)^{2}
\label{eq:chi2}    
\end{equation}
where $i$ is the data sampling point, $n$ is the total number of sampling points, the angular distance between sampling points is $0.1^{\circ}$. $S_{G}(\theta, E_{\gamma})$ is surface brightness predicted by the Gaussian profile, $S(\theta, E_{\gamma})$ is surface brightness predicted by the two-zone diffusion model, and  $\sigma_{G}^i$ is the  point error based on Gaussian profile error, which is adjusted according to the sample size $n$. The morphology derived from the two-zone diffusion model is also convolved with the same PSF. The free parameters in the fitting process are $q_0$ and $\alpha$ defined in equation~(\ref{eq:inj}), which determine the normalization and spectral shape of the gamma-ray emission, respectively.

As shown in Figure \ref{fig:Gaussian profile}, when the gamma-ray profile calculated by the two-zone diffusion model is consistent with the LHAASO measurement within the central zone, the profile at lager angular distance is significantly higher than the Gaussian template. This discrepancy arises because electrons spread widely when they escape from the slow-diffusion zone. Consequently, there are still many residual signals even after masking a region of $2.5\sigma$. Therefore, the leakage signals are expected to contribute a considerable portion to the DGE.

\section{Results} \label{sec:result}
\begin{figure}
   \centering
   \includegraphics[width=1.0\textwidth, angle=0]{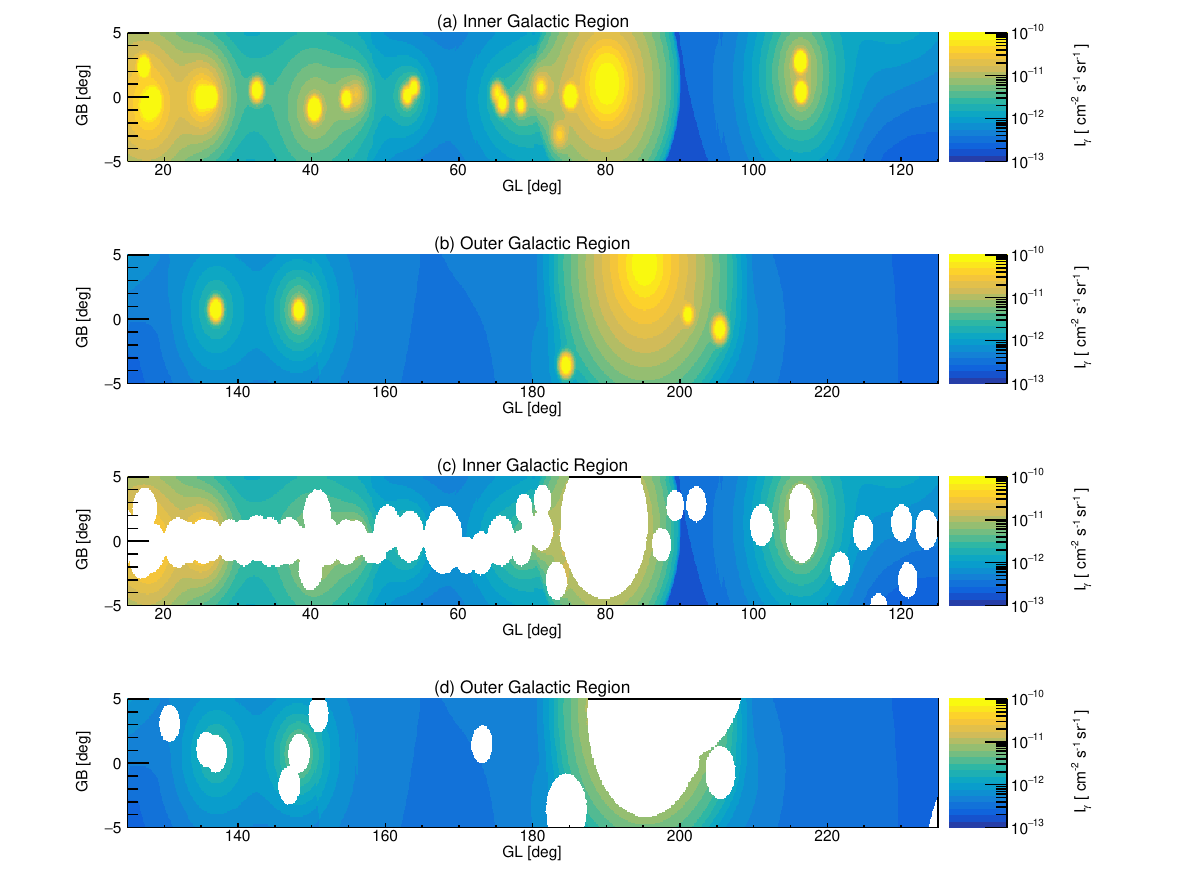}
   \caption{The gamma-ray intensity map of Galactic plane contributed by LHAASO PWNe/halos and Cygnus bubble in energy range 10-63 TeV. The signals from LHAASO PWNe/halos are modeled using a two-zone diffusion model. (a) The inner Galactic plane of $15^{\circ}<l<125^{\circ}$. (b) The outer Galactic plane of $125^{\circ}<l<235^{\circ}$. (c) The inner Galactic plane after masking the same region as the LHAASO DGE measurement. (d) The outer Galactic plane after masking the same region as the LHAASO DGE measurement. The white regions in (c) and (d) represent the masked regions.}
   \label{fig:source_map}
\end{figure}

Using the model given by the fit introduced in Section \ref{subsec:fit}, we present the original and masked gamma-ray intensity maps in the Galactic plane in Figure~\ref{fig:source_map}. The masking method employed is consistent with that used for the LHAASO DGE measurement. We can visually discern that considerable leakage signals remain in the Galactic plane, which should not be ignored compared to the measured DGE. In the default calculation, we assume $D_1 = 4.5 \times 10^{27}$~cm$^2$~s$^{-1}$ and $r_\star=25$~pc, as introduced in Section~\ref{subsec:2zone}. These two parameters cannot be constrained by the current measurements for each source, but they can affect the expected signal leakage. We discuss the influence of varying these parameters in Section~\ref{sec:discussion}.

It is worth noting that it is necessary to apply special treatments for the Cygnus and Geminga regions. In the region of Galactic longitude $70^\circ-90^\circ$, there is an ultrahigh-energy gamma-ray bubble powered by a super PeVatron which has an extension of $\sim 10^\circ$, dubbed the ``Cygnus bubble" \cite{LHAASO:2023uhj}. However, in the LHAASO measurement of DGE, only a $6^\circ$ radius circle is masked, leading to leakage of the bubble signals. The Cygnus bubble most likely originates from hadronic interactions \cite{LHAASO:2023uhj}, and we straightforwardly use the LHAASO measurement to estimate the leakage flux in the range of $6^\circ-10^\circ$ instead of using the prediction from the two-zone diffusion model. Since the hadronic emission generated by the ``cosmic-ray sea" is included in the LHAASO measurement of the Cygnus bubble, we have removed it to avoid redundant calculation.

At the Galactic longitude of around $190^\circ$, there is a remarkable excess coinciding with the location of the Geminga halo even after masking a radius of $8^\circ$ around the Geminga pulsar. For the Geminga halo, we assume a slow-diffusion zone with $D_1=1.5 \times 10^{28}$~cm$^2$~s$^{-1}$ and $r_\star=50$~pc. The predicted leakage signals can reproduce this excess, and the assumed parameters are reasonable for the Geminga halo. The extension of Geminga observed in the significance map reported by LHAASO-KM2A \cite{Guo:2021vuy} is larger compared to the measurements by HAWC \cite{Abeysekara2017}, indicating a larger $D_1$. Besides, \cite{Fang:2023xla} pointed out that $r_\star$ should be larger than $30$~pc to reproduce the surface brightness profile of the Geminga halo; otherwise, the profile would be too steep compared to the measurements.

\begin{figure}
   \centering
   \includegraphics[width=\textwidth, angle=0]{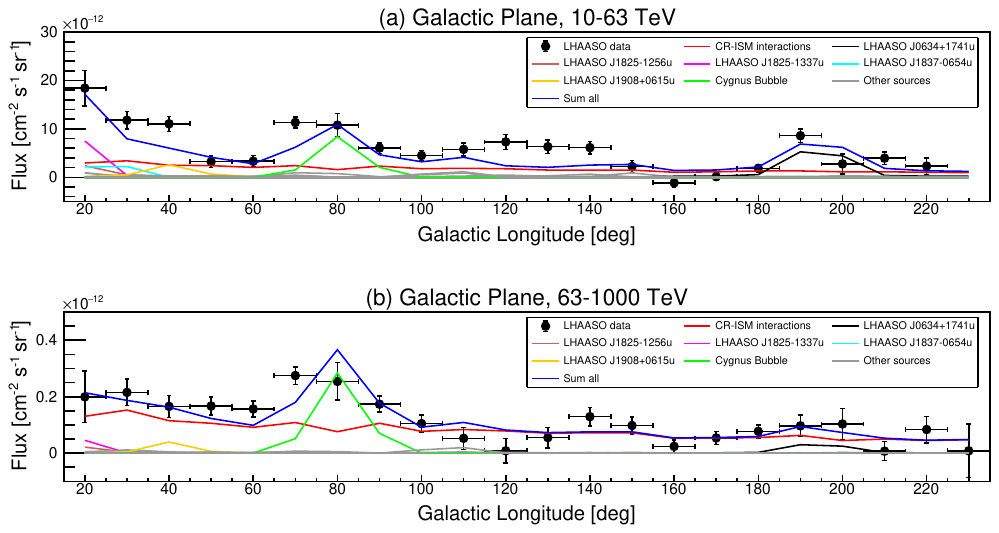}
  \caption{The comparison of Galactic profiles of the diffuse emission and the model. (a) In the energy range of $10-63$ TeV. (b) In the energy range of $63-1000$ TeV. The black points represent the measurements from LHAASO. The red line represents the prediction of the CR-ISM interaction model \cite{Yan:2023hpt}. The blue line represents the sum of contributions from the CR-ISM interaction and the leakage signals of the LHAASO PWNe/halos and the Cygnus bubble. For sources with significant signal leakage, we have specifically marked their contributions to the DGE.} 
   \label{fig:GL_component}
   \end{figure}
   
Figure \ref{fig:GL_component} shows the Galactic longitude profiles of DGE contributed by different components, including the CR-ISM interaction\cite{Yan:2023hpt}, and leakage signals of LHAASO PWNe/halos and Cygnus bubble, in the energy bands of $10-63$~TeV and $63-1000$~TeV. The DGE measurement results significantly exceed the expectations of CR-ISM interaction model. For the 10-63 TeV range, the DGE excess is primarily concentrated in four regions: (1) The region with $l\approx20^\circ - 50^\circ$; (2) Around the Cygnus bubble; (3) The region with $l\approx100^\circ - 150^\circ$; (4) Around Geminga. Our signal leakage model significantly contributes to the excesses in regions (1), (2), and (4). For region (3), the signal leakage model can partly explain the excess (in $l\approx100^\circ - 115^\circ$). For the 63-1000 TeV energy range, the excess primarily occurs in region (2), with a smaller amount coming from region (1). Our signal leakage model provides a good explanation for these excesses. 

Interestingly, the DGE excess around Geminga is primarily observed in the $10-63$~TeV energy range and is not significant in the $63-1000$~TeV range. This observation coincides with the characteristics of the Geminga halo spectrum: the latest measurement by HAWC indicates that the gamma-ray spectrum of the Geminga halo exhibits an exponential cutoff at tens of TeV \cite{HAWC:2023bfh}. These phenomena favor the idea that the excess in this region originates from leakage signals of the Geminga halo rather than other unresolved sources.

\begin{figure}[h]
  \begin{minipage}[]{0.5\linewidth}
  \centering
   \includegraphics[width=80mm]{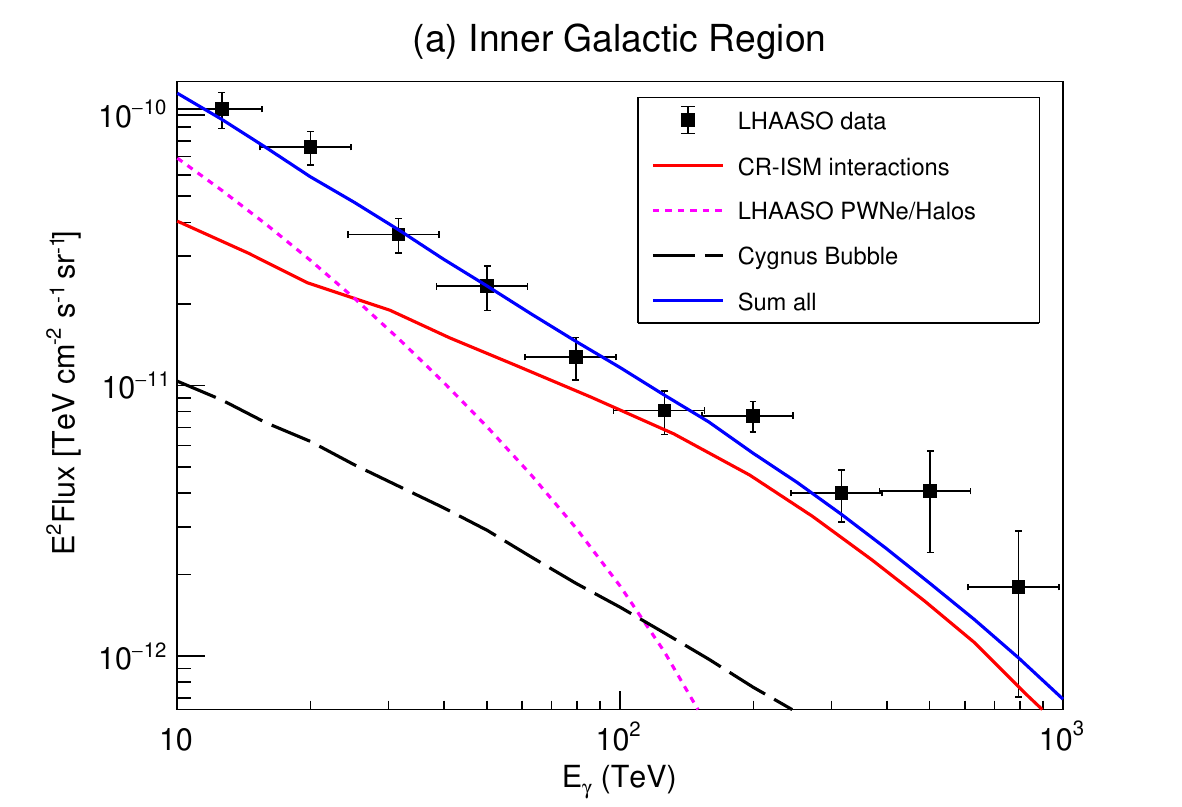}
  \end{minipage}%
  \begin{minipage}[]{0.5\textwidth}
  \centering
   \includegraphics[width=80mm]{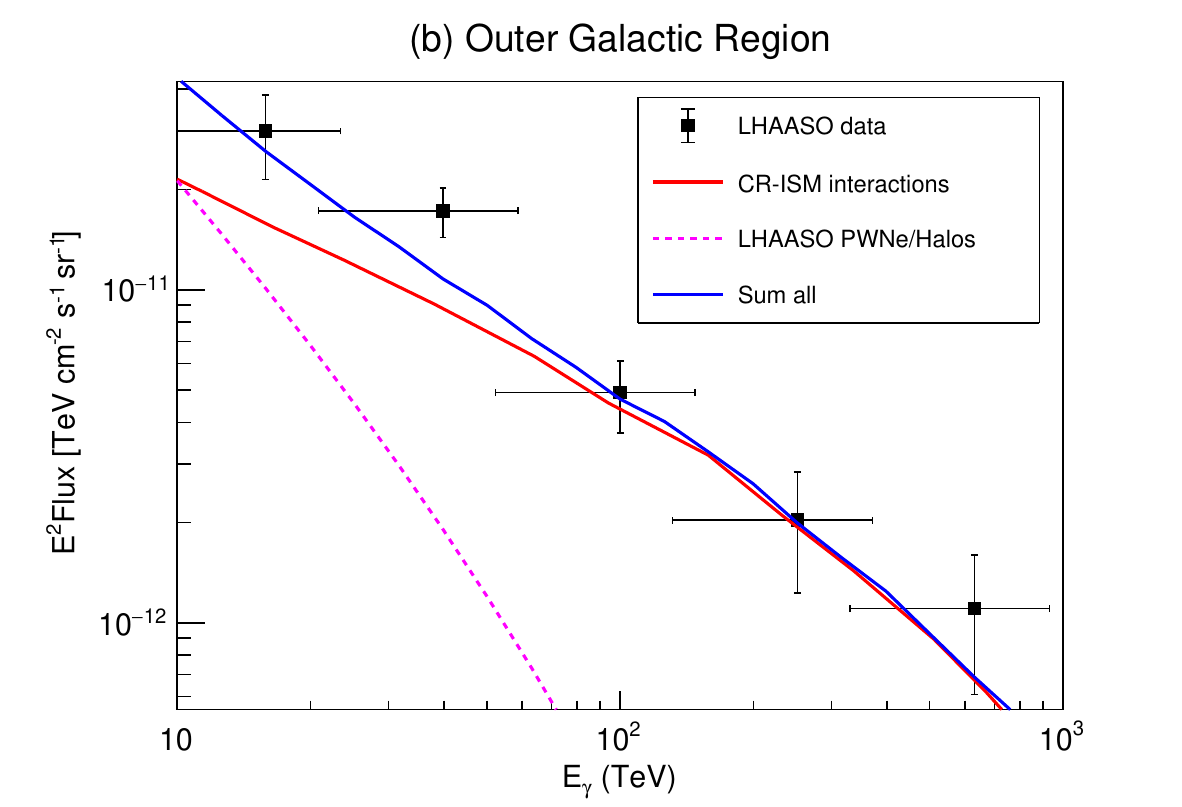}
  \end{minipage}%
  \caption{Comparison of the measured DGE spectrum and model interpretation. (a) In the inner Galactic plane of $15^{\circ}<l<125^{\circ}$. (b) In the the outer Galactic plane of $125^{\circ}<l<235^{\circ}$. For each panel, the black points represent the LHAASO measurement. The solid red line represents the prediction of the CR-ISM interaction model \cite{Yan:2023hpt}, the dotted pink line represents the contribution from the leakage fluxes of LHAASO PWNe/Halo, and the black dashed line represents the leakage flux of Cygnus bubble. The blue line is the sum of all these components.}
  \label{fig:sed_component}
\end{figure}

Figure \ref{fig:sed_component} shows the gamma-ray spectra of DGE contributed by different components in the inner and outer Galactic plane, respectively. The total flux of DGE can be well explained once the signal leakage components are taken into account. In the inner Galactic plane region, for energies below $100$~TeV, the DGE excess is predominantly contributed by the LHAASO PWNe/halos, while for energies above $100$~TeV, it is mainly contributed by the Cygnus bubble. This is consistent with the constraint from the neutrino observation, which indicates the DGE excess in $\approx1-100$~TeV is dominated by leptonic components \cite{Yan:2023hpt}. In the outer Galactic plane region, the DGE excess is contributed by leakage signals of LHAASO PWNe/halos, with no significant excess observed above $100$~TeV. We note that the model still underestimates the DGE flux at $\approx40$~TeV for the case of the outer Galactic plane. This discrepancy is due to the underestimation in the longitude range of $l\approx115^\circ-150^\circ$ as shown in the top panel of Figure~\ref{fig:GL_component}.

\section{Discussion} \label{sec:discussion}
The uncertainties of the model arise from the parameters of the electron injection spectrum and the parameters of electron propagation. The injection parameters ($q_0$ and $\alpha$) are constrained by the LHAASO spectral measurements of the sources. The propagation parameters are mainly constrained by the morphology measurements of the sources; however, LHAASO currently provides only measurements of Gaussian extents for most sources, resulting in weaker constraints on the propagation parameters. Therefore, the uncertainties of the model primarily stem from the propagation parameters, especially $D_1$ and $r_*$, which determine the amount of signal leakage.

\begin{figure}
   \centering
   \includegraphics[width=\textwidth, angle=0]{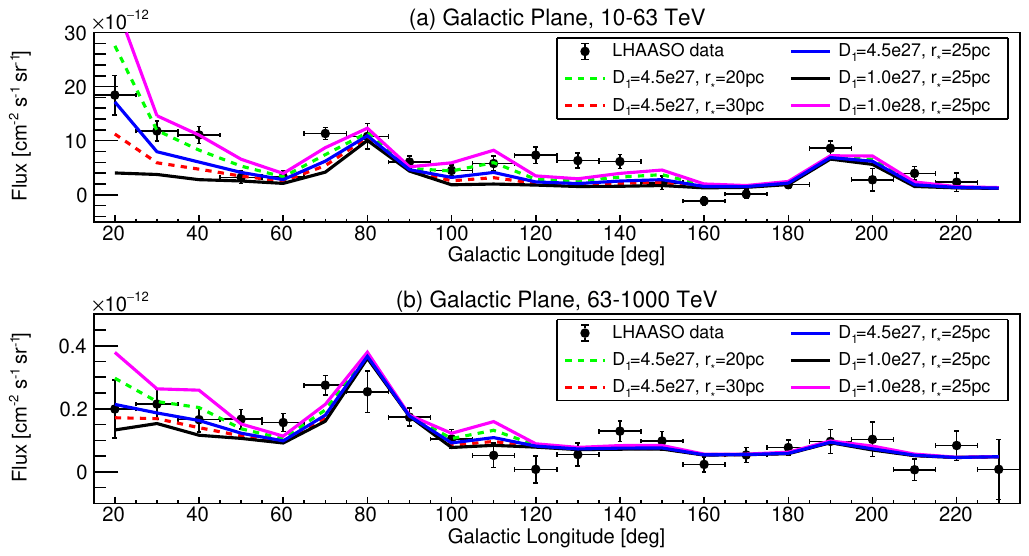}
   \caption{Comparison of the Galactic longitude profiles of DGE by assuming different two-zone diffusion model parameters. (a) In the energy range of $10-63$ TeV. (b) In the energy range of $63-1000$ TeV. The lines represent the total contribution from the CR-ISM interaction, the LHAASO PWNe/Halo, and the Cygnus bubble. The units of $D_1$ are cm$^2$~s$^{-1}$.}
   \label{fig:GL_multi}
   \end{figure}

In the two-zone diffusion model, the slow-diffusion zone around a pulsar may originate from the turbulent environment created by the associated SNR \cite{Fang:2019iym}. Given that most of the pulsars listed in Table~\ref{tab:pulsar} are on the order of tens of kyr, it is reasonable to assume an average slow-diffusion zone scale of $r_\star=25$~pc, which aligns with the expected scale of SNRs under typical parameters \cite{Leahy:2017nrs}. However, we still need to evaluate the impact of varying this value. Additionally, we should also consider the impact of the diffusion coefficient in the central zone, $D_1$, which may deviate from the value measured by HAWC \cite{Abeysekara2017}. Qualitatively speaking, the larger $D_1$ or smaller $r_\star$, the more easily electrons escape to distant regions, resulting in greater signal leakage. The specific impact of varying $D_1$ or $r_\star$ on the interpretation of the DGE excess is illustrated in Figure \ref{fig:GL_multi}.

We have demonstrated that the DGE excess in most regions can be interpreted by the signal leakage model under certain parameters, e.g., $D_1\approx5\times10^{27}$~cm$^2$~s$^{-1}$ and $r_\star\approx20-25$~pc. However, the gamma-ray flux predicted by the model is insufficient to match the data in the range of $l\approx115^\circ-150^\circ$, even when considering model uncertainties, as shown in Figure~\ref{fig:GL_multi}. There are relatively few LHAASO PWNe/halos in this region, and the fluxes of LHAASO J0249$+$6022 and LHAASO J0359$+$5406 located in this region are relatively low. The excess gamma-ray emission in this region may be caused by some unresolved leptonic sources, such as indistinguishable halos. Therefore, the signal leakage from known sources and the contribution from unresolved sources should be considered complementary in explaining the DGE excess.

In this work, we predict the large-angle signal leakage of LHAASO PWNe/halos using the two-zone diffusion model for electrons. In fact, besides the two-zone diffusion model, other propagation models may also explain the surface brightness distribution around pulsars while predicting significantly higher gamma-ray signals at large angles compared to the Gaussian template. The superdiffusion model exhibits such characteristics, and previous works have discussed using the superdiffusion model to explain the morphology of pulsar halos \cite{Wang:2021xph,Fang:2021qon}. We expect that such models can achieve similar results to the two-zone diffusion model in interpreting the DGE excess.

We note that some studies on the joint analysis of the LHAASO diffuse gamma-ray measurement and IceCube neutrino measurement suggest a hadronic origin for DGE \cite{Fang:2023ffx,Shao:2023aoi}. It is crucial to emphasize that a consistent treatment of both datasets is essential for a robust joint analysis \cite{Yan:2023hpt}. The neutrino flux measured by IceCube includes contributions from both the CR sea and hadronic point sources. To correlate the neutrino flux measured by IceCube with the diffuse gamma-ray data from LHAASO, the contributions from hadronic sources should be subtracted. Furthermore, the neutrino flux should be derived from the same region of interest as the LHAASO DGE analysis. If the same masking method used by LHAASO is not performed to the IceCube data, or if the contributions from hadronic point sources are not subtracted, it could lead to an overestimation of the neutrino flux, potentially resulting in the conclusion that the DGE excess has a hadronic origin.

\section{Conclusions} \label{sec:conclusion}
The DGE excess phenomenon in the Galactic plane is generally believed to originate from unresolved sources with large extensions, such as pulsar halos. In this work, we offer a new perspective to explain the DGE excess observed by LHAASO, suggesting that the large-angle signal leakage from known sources could be a significant contributor to the DGE excess. As the excess predominantly originates from gamma rays produced by leptonic processes, we focus on possible PWNe and pulsar halos in the LHAASO source catalog. We first use the two-zone diffusion model to fit the central profile of the sources and get the extrapolated profile at large angular distances. Then, using the same masking method as in the DGE measurement, we determine the contribution from leakage signals of the LHAASO PWNe/halos to the DGE.

We demonstrate that the DGE excess in most regions of the Galactic plane can be well interpreted by the signal leakage model under certain parameters. Additionally, we point out that signal leakage from the Cygnus bubble is also necessary to be considered. We also note that in the Galactic longitude range of $l\approx115^\circ-150^\circ$, the gamma-ray flux predicted by our model is insufficient to match the observation, as there are few identified bright gamma-ray sources in that area. Unresolved extended sources may account for this discrepancy. Therefore, we emphasize that the signal leakage from known sources and the contribution from unresolved sources should be considered complementary in explaining the DGE excess.

Besides the two-zone diffusion model, other propagation models, such as superdiffusion could also explain the surface brightness distribution around pulsars while predicting significantly higher gamma-ray signals at large angles compared to the Gaussian template, which may similarly account for the DGE excess. As the LHAASO data continues to accumulate and with the advent of future high-resolution observations (e.g., LACT \cite{Zhang:2023bsi}), we will be able to provide more precise measurements of the morphology of known gamma-ray sources. This will allow us to place more stringent constraints on complex propagation models, such as the two-zone diffusion and superdiffusion models, thereby offering better insights into the origin of the DGE excess.

\begin{acknowledgements}
This work is supported by the National Natural Science Foundation of China under Grants No. 12105292, No. 12175248, and No. 12393853. 

\end{acknowledgements}

\bibliography{main}

\end{document}